# COMPUTER-SUPPORTED COLLABORATIVE LEARNING IN SOFTWARE ENGINEERING EDUCATION: A SYSTEMATIC MAPPING STUDY

*Antti Knutas, Jouni Ikonen, Jari Porras*

Lappeenranta University of Technology
firstname.lastname@lut.fi
Finland

**Abstract:** Computer-supported collaborative learning (CSCL) has been a steady topic of research since the early 1990s, and the trend has continued to this date. The basic benefits of CSCL in the classroom have been established in many fields of education to improve especially student motivation and critical thinking. In this paper we present a systematic mapping study about the state of research of computer-supported collaborative learning in software engineering education. The mapping study examines published articles from 2003 to 2013 to find out how this field of science has progressed. Ongoing research topics in CSCL in software engineering education concern wider learning communities and the effectiveness of different collaborative approaches. We found that while the research establishes the benefits of CSCL in several different environments from local to global ones, these approaches are not always detailed and comparative enough to pinpoint which factors have enabled their success.

**Key words:** computer-supported collaborative learning, CSCL, software engineering education, systematic mapping study, SMS

## 1. INTRODUCTION

Collaborative learning, or cooperative activity of students working together towards a specific learning goal with the teacher as a facilitator [10, 25, 50], has become an increasingly important topic in education [75]. This collaborative approach to education has been shown to develop critical thinking, deepen the level of understanding, and increase shared understanding of the material [39, 50, 51, 95]. Computer-supported collaborative learning (CSCL) facilitates this collaboration by using computer-mediated communication tools to either enable new communication methods between students or to extend the range of communication beyond a single classroom [7, 24, 55, 82].

When citing, please use the following information: Knutas, A., Ikonen, J., & Porras, J. (2015). COMPUTER-SUPPORTED COLLABORATIVE LEARNING IN SOFTWARE ENGINEERING EDUCATION: A SYSTEMATIC MAPPING STUDY. International Journal on Information Technologies & Security, 7(4).

The extension of collaboration with CSCL allows increased knowledge building between a wider range of participants, more flexible teaching structures independent of place or time, better monitoring of student understanding by instructors, and improved student productivity and satisfaction [82]. However, Williams and Roberts [106] point out that the nature of CSCL has to be taken into account from the first planning stages when designing courses and it has to be clearly explained to the students. If not implemented properly, poorly designed CSCL will be a drawback instead of a benefit [106].

Computer-supported collaboration is essential in software engineering education, because working and efficiently collaborating teams are at the basis of the software engineering industry [18]. Additionally, recent trends in software engineering head towards continuous computer-supported collaboration within teams [45, 73]. These methods are common in the industry and the use of these techniques in higher education is increasing [83]. Common computer-supported collaborative techniques in software engineering education include the use of code repositories and project management tools, especially in project management and capstone –style courses [15, 72]. In more lecture-oriented courses using virtual learning environment communication platforms to engage student have been shown to increase engagement and learning outcomes [29]. This is because the *information revolution in the classroom not only makes more information available to the student, but it also enhances interactivity and provides new collaboration channels* [38].

The next subsections describe our research goals and examine earlier literature studies in the field. Section 2 explains the steps taken in the mapping study. The results of the mapping study are presented in section 3. Section 4 discusses the research trends of CSCL in software engineering education, and section 5 concludes the study.

### 1.1. Research goals

Computer-supported collaborative software engineering education includes different levels and different types of collaboration [18]. The collaborators range from local students [70] to globally cooperating learning networks [36]. In this study, the research goal is to *discover the extent of collaboration in computer-supported collaborative learning as used in software engineering education,* and specifically *the range of collaboration*. To achieve this goal, we have examined earlier studies conducted in the field systematically and arranged them by the distance and variety of entities engaging in communication. The specific questions we set for this systematic literature review study are:

1. What have been the publication trends in studies about computer-supported collaborative learning in systematic literature reviews in software engineering education between 2003 and 2013?

2. What aspects and ranges of collaboration have been examined in these studies?
3. What research methods have been used?
4. Are there research gaps in the field of study or areas of computer-supported collaboration that could still be studied further?

### 1.2. Earlier literature reviews on computer-supported collaborative learning in software engineering education

We were not able to identify literature reviews that examined the issue of computer-supported collaborative learning in software engineering education directly. However, there were several literature reviews about CSCL in general and other fields of CSCL that touched the issue of software engineering education indirectly. These literature reviews included a review of the technologies used in CSCL, an overview of case studies about CSCL, and a review that collected and established reporting standards for the field of CSCL.

The different technologies used in CSCL have been studied by Resta and Lafarrière [82] in their literature review "Technology in Support of Collaborative Learning." They review the recent trends in CSCL research and the beneficial outcomes for CSCL, and identify instructional motives for using CSCL. Additionally, their review includes several recommendations for the directions of research, including recommendations to investigate the unique benefits of CSCL instead of comparing CSCL to face-to-face learning and to investigate the organizational requirements for arranging CSCL.

Two literature reviews released in the 2000s concern case studies published in the field of CSCL. "Instructional Methods for CSCL: Review of Case Studies" published by So and Kim [96] reviews ten cases in order to identify the instructional goals, methods, effectiveness and conditions of online collaborative learning. Hammond [41] reviews recent publications in his article "A Review of Recent Papers on Online Discussion in Teaching and Learning in Higher Education" and examines the publications' curriculum design, assumptions about teaching, and reported conditions for using online discussion. Both articles emphasize the importance of the need to develop curriculum models, the need of proper support by an instructor, and the impact of the software environment on communications.

The reporting standards for CSCL have been collected and established by Hadwin et al. [40] in their article "Toward Standards for Reporting Research: A review of the literature on Computer-Supported Collaborative Learning." They have reviewed articles on methodologies, theoretical and operational definitions and collaborative models and found out that many publications use diverse reporting methodologies and terminology. Hadwin et al. [40] attribute this to the cross-discipline nature of CSCL and propose standards for reporting on collaborative models, tools and research.

## 2. RESEARCH METHOD

A systematic mapping study (SMS) is a secondary study that aims at classification and thematic analysis of earlier research [56, 78]. It is closely related to a wider secondary study, a systematic literature review (SLR), which aims at gathering and evaluating all the research results on a selected research topic [2, 57]. Kitchenham and Charters [56] present the best practices of both for the field of software engineering and also compare the two. The SMS is more general in search terms and aims at classifying and structuring the field of research, while the target of SLR is to summarize and evaluate the research results. Kitchenham and Charters [56] also discuss the applications and states where SMS can be especially suitable if few literature reviews have been done on the topic and there is a need to get a general overview of the field of interest. Both kinds of studies can be used to identify research gaps in the current state of research.

A systematic mapping study classifies and structures a field of interest in research by categorizing publications and analyzing their publication trends [78]. Additionally, SMS can analyze what kind of studies have been done in the field, and what are the research methods and outcomes [5]. In Table 1 we present how we have used the systematic mapping study method created by Bailey et al. [5] for the field of software engineering, developed further by Petersen et al. [78].

*Table 1. SMS search procedure steps*

| Step | Procedures | Results |
| --- | --- | --- |
| 1. Determine search terms. | Determine search terms from accepted field keywords that cover the desired topics. | Boolean search terms that get desired results from the databases in the next steps. |
| 2. Determine databases. | List the databases that cover most of the publications in the chosen field of science. | A list of databases for steps three to four. |
| 3. Run a test search. | Select one database and run a search to test the validity of the search terms. | Verification that the search terms will return the desired type of publications. |
| 4. Run a full search and store the results. | Search all the selected databases and store the results. | A list of all publications that match the search terms. |
| 5. Deduplicate and sort according to inclusion and exclusion criteria. | Remove duplicate results and then use the inclusion and exclusion criteria to select the articles for in-depth analysis. | The final list of articles that will be included In the systematic map. |
| 6. Analyze the query results. | Review the articles and record the determined data and metadata. Analyze and compare the research articles and their research approaches. | A systematic map of the chosen section of research literature. |

## 2.1. Keyword and database selection

We started the article search process for our study by doing first a survey about current keywords and research trends in computer-supported collaboration in software engineering education. This survey was sent to 131 recipients chosen from the editorial boards and authors in publications related to computers in education or computer-supported collaborative research. The results of this survey were examined and used in the pilot search process. However, the results of these pilot searches were too narrow when tested with the suggested keywords.

To address the issue of too few articles found, we examined recent publications in the field of CSCL and the most commonly used keywords in them, and tested whether these keywords were hypernyms of the keywords suggested in the survey results. Two keywords stood out: computer-supported collaborative learning and software engineering education. A final pilot search with the Boolean parameters of (("cscl" OR "computer-supported collaborative learning") AND "software engineering") yielded a number of results between a dozen and slightly over a hundred, depending on the database. This number of results was large enough to conduct a systematic mapping study.

The survey suggested several publications to include in the SMS. Instead of choosing individual publications for the search, we chose seven databases that included most of the publications in the field of software engineering education, computer-supported collaborative learning, and several closely related fields like education in general and software engineering. Another criterion for choosing these databases was the support for the Boolean operators present in the search terms and filtering functionality that enabled the exclusion of unrelated fields. The pilot searches emphasized the necessity for the filtering feature, because CsCl, or cesium chloride, is a common term appearing in chemistry publications.

The search terms for the actual search were the same as in the last pilot search with the year range of 2003 – 2013. The searched databases, the search criteria, the total number of results and the number of papers included in the study after the application of the inclusion and exclusion criteria are listed in Table 2. All the searches were done with the same Boolean search parameters of (("cscl" OR "computer-supported collaborative learning") AND "software engineering"). In some databases several searches were performed, because they did not support searching from metadata and abstracts at the same time. The search parameters included both conference and journal publications when available.

*Table 2. Summary of database query results*

| Source database | Search parameters and Filtering | Included / Total |
|---|---|---|
| EBSCO | 1) Peer-reviewed journals between 2003 and 2013, search from titles.<br>2) Peer-reviewed journals between 2003 and 2013, search from abstracts. | 19 / 120 |
| IEEE Xplore | Journals and conference publications between 2003 and 2013, search from metadata and abstracts. | 23 / 23 |
| Springer Link | Journals between 2003 and 2013, search from metadata and abstracts. Exclude journals related to physics and materials sciences. | 13 / 67 |
| Science Direct | Journals between 2003 and 2013, search from metadata and abstracts. Exclude journals related to physics and materials sciences. | 24 / 88 |
| ACM Digital Library | Journals and conference publications between 2003 and 2013, search from metadata and abstracts. | 51 / 134 |
| CiteSeer | Journals between 2003 and 2013, search from metadata and abstracts. | 2 / 2 |
| Emerald Insight | Journals between 2003 and 2013, search from metadata and abstracts. | 3 / 7 |
| Total | | 121 / 433 |

### 2.2. Analyzing the query results

A total of 433 conference and journal articles were found in the database searches. They were first reviewed by reading the title, keywords and abstract. In the first round of review, articles that did not in any way discuss computer-supported collaborative learning or software engineering education, or were written in other languages than English were dropped from the study. After the first round of reviews and article deduplication, 121 articles were selected for an in-depth review and comparison against the inclusion and exclusion criteria.

#### 2.2.1. Inclusion and exclusion criteria

During the second round of review, the inclusion and exclusion criteria were applied to the remaining articles. The inclusion criteria in this study were discussion of the following topics:

- social networking or collaboration in software engineering education,
- social networking or collaboration in intensive or team/project-based education, or
- use of CSCL in software engineering education.

The excluded categories in the papers were:
- literature surveys with no original research,
- papers not subject to peer review, or
- papers not considering the research topic from the perspective of CSCL, collaborative, engineering or computer science education.

If a paper discussed CSCL and only tangentially touched the inclusion criteria, it was still included in the SLR study in order to give as comprehensive a view of research as possible. After this final round of filtering, a total of 78 articles were selected to be included in the systematic literature review.

### 2.2.2. Article categorization

The articles were categorized according to the research goals of the study, which were to map the range and diversity of collaboration in CSCL in SWE. Both content data and metadata were recorded from the articles, and each category of data is presented in Table 3 below.

*Table 3. Paper metadata and content data collection categories*

| Article metadata | Article content data |
| --- | --- |
| *Publication Name* <br> *Publication Type* (conference or journal) <br> *Publication Forum* <br> *Publication Year* (2003 – 2013) <br> *Keywords* | *Range of Collaboration* (general, single class, virtual learning environment, learning community, inter-community) <br> *Main Issue* (CSCL tool development, introducing a CSCL method, implementing CSCL, pedagogical or behavioral issues, effects of CSCL, developing research methodology) *Research Method* (constructive, literature study, case study, multiple case study) |

The first main category that was used to divide the papers was the different ranges of collaboration, i.e. general, single class, virtual learning environment (VLE), learning community or inter-community. These subcategories were chosen to map and emphasize the different scales of collaboration. If the research had applications on more than one range of collaboration, the wider one was noted. General articles did not specify the intended use of collaboration or discussed the field in general, single class articles studied collaboration that occurred mostly in a single classroom or a single course, and VLE articles studied collaboration in distance education or in situations where the participants were not physically present. In the last two subcategories there were one or several communities involved, which surpassed the context of a single course in scale or duration.

The second main category for dividing the articles was the main issue discussed in the article, which were tool development, introducing or implementing a CSCL method, pedagogical or behavioral issues, effects of CSCL, and developing a research methodology. These categories were created inductively by reviewing the articles and labeling the main topics of research and most common article themes.

The research method was not always explicitly mentioned in the article. If the research article discussed earlier literature or proposed a new theory based on earlier literature, the research method was marked to be a literature study. If the paper used a constructive or a design science approach [44], in which the research artifact was an implemented software system or a new testable teaching plan, the method was marked to be constructive.

### 3. MAPPING THE STUDY RESULTS

On the basis of the search results, it can be said that the interest in the field has been steady with some variations in yearly publications. In the early 2005 the field of research had a spike of publications with steady decline until 2008, after which the number of publications has been rising slowly. Figure 1 shows the number of publications included in this study year by year. Several of the excluded papers discussed CSCL in general without mentioning software engineering education, 14 in the year 2008, so the interest in the more general field of CSCL seems to be strong.

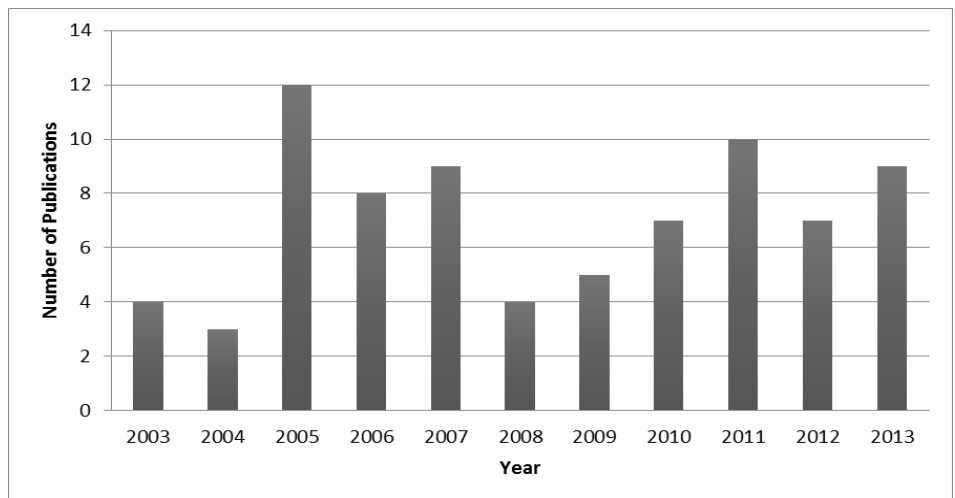

*Fig. 1. The number of mapped publications by year*

The most commonly addressed issue was the development and introduction of new CSCL tools, with 20 articles discussing this area. These articles often took a constructive approach to the problem of collaboration, pinpointing some issue in collaboration and introducing a new tool to facilitate collaboration. These articles also often included a case study to validate the intended effect of the tool and to show that it facilitates collaboration in the intender manner. Closely related but less numerous were the articles introducing a new CSCL method, with 8 articles addressing this issue. They concentrated less on the facilitating software tool and introduced instead a new teaching or organizational method for implementing computer-supported collaboration.

Almost as numerous were the articles that detailed the implementation of CSCL in some environment, with 15 articles addressing this issue. They were often single or multiple case studies that described the environment or organizations, the chosen CSCL approach, and how they implemented it. Compared to the nine articles that discussed the effects of CSCL, they concentrated more on how to implement a more in-depth study of the effect and implications on learning outcomes.

Sixteen articles discussed pedagogical or behavioral issues. They either developed a new pedagogical theory for computer-supported collaborative learning or presented a study about how CSCL affects the collaborative behavioral patterns of students. Compared to articles that studied the effects of CSCL, the behavioral studies emphasized the examination of social behavior instead of learning outcomes.

Lastly, there were 8 articles concerning the development of research methodologies for studying, data collecting, or assessing computer-supported collaborative learning. These articles adapted for example a previous research method, like interaction analysis, and presented how it can be applied to CSCL or establish guidelines for assessing different aspects of CSCL, like student collaboration.

Table 4 lists the research methods used in the articles reviewed in this study. The most noticeable trend is that the most common research approach is constructive, and many of the papers introduced something new and then evaluated it. The second most common approach is case studies, which is logical when considering the educational setting: many of the papers evaluated the success of a certain approach in the context of a course, where each course can be considered a test case. A minority of the papers with this research approach were multiple case studies, which means that in many of these papers the study concerned a single test environment only.

*Table 4. Research methods used in the articles*

| Research method | Number of publications |
|---|---|
| Case Study (CS) | 24 |
| Multiple Case Study (MCS) | 8 |
| Constructive Research (CR) | 40 |
| Discussion Paper (DIS) | 1 |
| Survey (SUR) | 1 |
| Literature Study (LIT) | 4 |

A systematic map of computer-supported collaboration in software engineering education is presented in Table 5. In this map the articles are arranged in the horizontal axis according to the scope of collaboration evaluated in the study, and according to the main research issue in the vertical axis. The next subsections go into more detail about the state of research in each specific issue, present the main findings in the articles, and interpret the systematic map.

*Table 5. A systematic map of collaborative learning in software engineering education*

|  | General *(9 articles)* | Single class *(29 articles)* | Virtual learning environment *(17 articles)* | Learning communities *(21 articles, 10 single community, 11 multiple community)* |
|---|---|---|---|---|
| CSCL Tool Development *(20 articles)* | Pansanato & Fortes 2005 [76], CR<br>Kahrimanis et al. 2006 [52], CR<br>Vega-Gorgojo et al. 2006 [101], CR<br>Liu & Wang 2012 [63], CR | Hübscher-Younger & Narayanan 2003 [48], CR<br>Carroll & Rosson 2005 [14], CR<br>Coelho & Murphy 2007 [19], CR<br>Baghaei et al. 2007 [4], CR<br>Yuan & Jin 2008 [107], CR<br>Milentijevic et al. 2008 [72], CR<br>Chen & Teng 2011 [15], CR<br>Kilamo et al. 2012 [54], CR<br>Cabrera-Lozoya et al. 2012 [12], CS | Lonchamp 2005 [64], CR<br>Phelps et al. 2005 [79], CR<br>Feng et al. 2006 [30], CR<br>Bravo et al. 2013 [9], CR<br>Sampaio et al. 2013 [86], CR | *Single Community:*<br>Cobos & Pifarré 2008 [17], CR<br>Rubens, Vilenius, & Okamoto 2009 [85], CR |
| Introducing a CSCL Method *(8 articles)* | Nickel & Barnes 2010 [74], DIS | Maresca et al. 2011 [67], CR<br>Tsompanoudi et al. 2013 [100], CR |  | *Single Community:*<br>Sancho-Thomas et al. |

| | | | | 2009 [87], CR<br>Repenning et al. 2009 [81], CR<br><br>*Inter-Community:*<br>Rohde et al. 2005 [84], CR<br>Collazos et al. 2010 [20], CR<br>Giraldo et al. 2011 [37], CR |
|---|---|---|---|---|
| Implementing CSCL<br>*(15 articles)* | | LeJeune 2003 [60], CS<br>Hübscher-Younger & Narayanan 2003 [47], CR<br><br>Cubranic et al. 2006 [22], CS<br>Hernández-Leo et al. 2007 [43], CS<br>Chou & Min 2009 [16], CS | (Schümmer et al. 2005 [90], CR<br><br>Basawapatna & Repenning 2010 [6], CR<br>Caballé et al. 2011 [11], CR<br>I. O. Elmahadi & Osman 2012 [28], SUR | *Single Community:*<br>Sheard 2004 [94], CS<br><br>Fischer et al. 2007 [32], MCS<br>Papadopoulos et al. 2013 [77], CS<br><br>*Inter-Community:*<br>Serce et al. 2010 [91], MCS<br>Giraldo et al. 2010 [36], MCS<br>Coccoli et al. 2011 [18], CR |
| Pedagogical or Behavioral Issues<br>*(16 articles)* | | Asensio et al. 2004 [3], CR<br>Dunlap 2005 [26], CS<br>Alfonseca et al. 2006 [1], CS<br>Kokubo et al. 2007 [59], CS<br>Karakostas & Demetriadis 2011 [53], CR | Harrer et al. 2005 [42], CS<br>Vivian et al. 2013 [102], CS | *Single Community:*<br>Knutas et al. 2013 [58], MCS<br><br>*Inter-Community:*<br>Carroll & Farooq 2007 |
| | | Warin et al. 2011 [104], CR<br>Wang & Hwang 2012 [103], CS<br>Ferreira 2013 [31], CS | | [13], CR<br>Liccardi et al. 2007 [61], CS<br>Serçe et al. 2011 [92], CS<br>Swigger et al. 2012 [97], MCS<br>Swigger et al. 2012 [98], CS |
| Effects of CSCL<br>*(9 articles)* | | Scheele et al. 2005 [88], MCS<br>Martinez-Mones et al. 2005 [70], MCS | Ghislandi & Job 2005 [35], CS<br>Dewiyanti et al. 2007 [23], CS<br>Glassman & Kang 2011 [38], LIT<br>I. Elmahadi & Osman 2013 [29], CS<br>Shaw 2013 [93], CS | *Single Community:*<br>Schellens & Valcke 2006 [89], CS<br>Ge et al. 2006 [34], CS |
| Research Methodology<br>*(8 articles)* | Martínez et al. 2003 [68], CR<br>Martínez et al. 2004 [71], LIT<br>Martinez-Mones et al. 2008 [69], LIT<br>Duque et al. 2009 [27], CR | Marcos et al. 2006 [66], CS<br>Marcos-García et al. 2007 [65], CR<br>Borge & Carroll 2010 [8], CR | Pimentel et al. 2005 [80], CR | |

### 3.1. Introducing a new CSCL tool or method

There were two main categories that stood out when reviewing articles that introduced novel approaches to CSCL. These articles concentrated either on a new computer-supported collaborative learning method and how to apply it, like a distributed learning approach by Collazos et al. [20], or a new CSCL software tool like the one by Chen and Teng [15] for managing student software projects.

The most numerous single category of articles was the one that introduced new CSCL tools, and it could be sorted into subcategories based on the scale of collaboration they facilitated. An exception to this were several papers evaluating the design or interoperability of CSCL tools at a more general level. The articles that addressed design considered CSCL tools from the perspective of metadata generation [76] and analysis and modeling [63]. Additionally, there were articles that addressed interoperability [52] and service discovery [101] in collaborative tools.

There were several articles describing the implementation of collaborative tools with a constructive approach to be used in the context of a single course or classroom. These articles discussed the collaborative learning of algorithms [48], a case library of teaching usability engineering [14], collaborative problem-solving in CSCL environments [4], combining several learning modalities [12], and an intelligent whiteboard teaching system [107]. One article took a different approach for collaboration and introduced a mentoring and peer reviewing system for classes [19]. There were two studies that worked in the context of a single course, but took a wider project- and team-based approach to collaboration. They examined the use of version control [72] and providing a platform for distributed development of student software projects [15].

The second subcategory of papers about software tools concentrated on virtual learning environments that support collaboration. The context of these was still a single class or a course, but they supported distance working. These studies about VLE could be divided into two different topics of communication support [30, 64] and programming groupware [9, 79]. Sampaio et al. [86] took a different approach in allowing students to challenge each other by building quizzes for course material collaboratively.

There were two articles that were directly about building tools for computer-supported collaborative learning communities. The first one introduced a system for collaborative knowledge construction in the web [17], and the second article introduced a tool for automatic group formation of community members [85].

The articles that introduced new CSCL methods could be divided to articles that either suggested new methods of organizational cooperation or proposed a novel method of applying CSCL to a learning environment. Rohde et al. [84] made a case for cooperation with external organizations in order to build communities practice. A similar project was introduced by Maresca et al. [67], where students worked on a large-scale open-source software project among professionals.

Several articles introduced methods for cultural changes for cooperation, by making the collaboration more agile and informal [74, 81] or by introducing a gamified student-centric reputation system [54]. More learning frameworks that emphasized acquiring collaborative teamwork skills were introduced with a problem-based approach [87], a distributed holistic approach [105] and a scripted approach [100]. Giraldo et al. [37] expanded on distributed learning environments and introduced a collaborative and distributed learning activity, as well as presented a case study where students from six different universities participated in a software project with positive project outcomes.

*There was a wealth of CSCL tools and methods introduced in the field over the time period examined in the SMS. These ranged from computer or communication tools improving or supporting certain aspects in collaboration to full pedagogical approaches, like problem-based learning. The papers introduced new, beneficial computer-supported tools or working methodologies in the classroom or communities. However, the papers concentrating on introducing new tools usually addressed a single use case or a communication problem and did not address the wider problem of interoperability with other tools. Only two papers considered the issue of interoperability directly, and later articles did not take these issues into consideration.*

### 3.2. Implementing CSCL

The second most numerous category was articles considering the implementation of CSCL in some setting. Instead of introducing a completely novel way of implementation, these articles took a method or tool and presented case studies, comparisons, evaluations or guidelines for the implementation of the method. These papers contained diverse approaches to evaluating or presenting CSCL implementation, but they could also be divided into different ranges of collaboration.

Several articles detailed different approaches and studies that examined the implementation of CSCL in the context of a single class. Cubranic et al. [22] compared different communication methods for novice programmers, concluding that instant messaging chat and video screen sharing worked best in their case. Chou and Min [16] studied the role of media presented in virtual CSCL environments. LeJeune [60], by contrast, studied the common elements of courses using CSCL, and concluded that there are several critical components in courses, including common task, collaborative behavior, positive interdependence, and both individual *and* group accountability and responsibility. Lastly, two studies discussed specific case studies. They explained how they implemented a CSCL scenario using open source software tools [43] and a combination of physical and online collaborative environments [6].

Some studies concentrated on experiences of virtual learning environments. The first such study included in this review was a report by Schümmer et al. [90] about a blended learning environment for teaching distributed software development. Another study concerned implementing a virtual environment - supported collaborative course in a culture where face-to-face collaboration is the common approach, and reported positive survey feedback from the participating students [28]. Two articles took a closer view at pedagogical roles in online environments, investigating the roles of people participating in discussions. The first article [99] examined the role of the teacher in online collaborative learning, and the second one presented practices for feedback, monitoring and evaluation in online collaborative discussions [11].

Several studies about CSCL implementation examined cases where one or several communities used the same collaboration system. The first article found in the search presented strategies for establishment and management, concluding that the establishment process is mostly student-driven [94]. A later article by Serce et al. [91] presented strategies and guidelines for building effective globally distributed student learning teams. Several articles presented case studies and positive experiences about distributed learning communities in software engineering education in Italy [18], Europe [77] and Latin America [36]. An article by Fischer et al. [32] proposed that community-based learning could be a strength of research-based universities and explained how the approach could be integrated to the computer science curriculum.

*Several articles considered the implementation of CSCL in local and global contexts, with overall positive results. These studies reported that computer-supported collaboration can be established either locally or in globally distributed teams, and that initial surveys had positive results. However, the wider the studies were, the less in-depth ones they generally were. When comparing the papers in this section with the ones considering the effects of CSCL in section 3.4., these papers covered more ambitious case studies at the expense of depth. Overall, the studies showing positive experiences of wide collaborative communities were promising. The basic premise of wide computer-supported collaborative communities was shown to be valid, and in the future in-depth studies could analyze the success factors of different CSCL approaches.*

### 3.3. Pedagogical and behavioural issues

Many articles discussed implementation details or presented detailed studies about computer-supported collaborative learning, but had a distinct focus that was not directly about collaboration or the tools used. Instead, they considered CSCL courses from a pedagogical or a behavioral point of view. Additionally, two articles considered the role of social networks [61] and task cohesion in regard to efficiency [103] in CSCL.

The articles that studied different pedagogical issues in CSCL were diverse in their approaches and topics. Three separate articles considered how students can acquire professional, creative and cognitive skills [26, 31, 104]. One article considered the impact of learning styles on student grouping [1], and another the effect of communication support on learning scenarios [42].

Several studies were conducted about how students collaborate and behave in CSCL learning scenarios. The studies were diverse, with several different ranges of collaboration covered from individual students [102] and groups [59] to globally distributed student teams [97]. As regards the topic, behavioral analysis can be divided to studies supporting tool or learning pattern design, collaborative communication patterns, and the behavioral patterns of distributed student teams.

Articles that concentrated on generalizing learning patterns developed for community-based learning [13], presented adaptation patterns for the operation of CSCL systems [53] and studied the convergence of authority in collaborative learning environments [47]. Additionally, Asensio et al. [3] presented how these learning patterns can be used to assist the development of component-based CSCL systems.

There were numerous articles that studied collaborative patterns, and these studies concentrated either on team behavioral patterns or collaborative communication patterns. Serçe et al. [92] and Swigger et al. [97, 98] presented findings on the behavior of global, distributed student teams. Their results showed that the communication patterns were related to the teams' communication modes, the nature of the task and the experience level of the leader, and that there was a positive correlation between the communication patterns and project outcomes. Knutas et al. [58] and Vivian et al. [102] studied the communication patterns that occur in CSCL classes and identified patterns and teamwork roles that emerge during the progress of the course.

*Several studies investigated how the implementation of CSCL affects classroom communication patterns and students' collaborative behavior. The difference from other articles considering the impact of CSCL was that these studies considered CSCL from pedagogical or behavioral points of view. The studies showed that students' behavioral and methods of organization affect the effectiveness of CSCL, and that properly used, CSCL can have a positive impact even on globally distributed learning communities. Additionally, the studies introduced in this section can be used in the design of new CSCL learning approaches or tools.*

### 3.4. The Effects of CSCL

Several articles concentrated on case studies that researched or presented the effect or impact of CSCL on learning scenarios or learning outcomes. These articles were often similar to the ones that considered the implementation of CSCL, but concentrated less on the implementation details and more on the effect. The studies included in this review presented results of the impact of CSCL on course communications, the dynamics of learning communities, knowledge construction processes, and the outcomes of problem-based learning. More specifically, the effects of CSCL were studied in interactive lectures [88], course-centered online communities [70], virtual learning environments [23, 29, 35, 38, 93], and collaborative online communities [34, 89].

In several studies, CSCL tools had affected course-centered communications positively in both online and physical environments. Using CSCL tools to make lectures interactive in a case study [88] was found increase student engagement and interest and enhanced learning. Another case on online communities [70] found that students learn concepts better on their own, but they are able to generalize collaboratively discussed concepts better.

Many studies had positive experiences of virtual learning environments. Using VLE communication platforms to collaborate around a course increased student satisfaction and motivation to work collaboratively, and had a positive effect on group process regulation and cohesion [23, 35]. Glassman and Kang [38] pointed out that one of the enabling factors in the information revolution in the classroom is not the amount of information available online, but the interactivity and new collaborative channels it provides. In VLEs, collaboration was shown to increase the learning outcomes [29], especially in small groups with external support [93]. Schellens and Valcke [89] had similar results, where collaborative knowledge construction in groups increased cognitive interaction and task orientation, with group size being a major variable, and smaller groups faring better. A paper by Ge et al. [34] proposed that successful open source software development community projects should be investigated and their online knowledge construction processes adapted to improve collaborative learning methods.

*The effects of CSCL have been studied widely and overall positive experiences from shared knowledge construction and student motivation to learning outcomes have been reported. However, most of the case studies presented in this systematic mapping study concentrated on course-centered communities and virtual learning environments. The question is whether CSCL and computer-supported collaboration communities can have as clear and specific benefits when implemented at a global scale, as proposed by Ge et al. [34] and presented in several case studies [18, 20, 37, 77]. The impact of these wider collaboration approaches should be studied further and analyzed comparatively in order to establish the benefits of CSCL at all levels of collaboration.*

### 3.5. CSCL research methodologies

The articles that developed research methods for the study of computer-supported collaborative learning were fewest in number, but perhaps most important in significance. With CSCL being a multidisciplinary field, several articles combined existing methods and presented how these can be applied in the field of CSCL. One of the basic approaches in collaboration analysis, social network analysis combined with qualitative evaluation, was presented by Martínez et al. [68]. This approach is in common use, including several studies presented in this review. It was further developed with approaches for automatic data collection in CSCL systems by Martínez et al. [71] and other research teams [27].

Collaboration analysis has also been used as a means for evaluating student performance in classrooms [8, 80]. This approach has been said to be critical, because standard student evaluation approaches are no longer sufficient in the collaborative classroom [80].

*The field of CSCL research has established research methodologies that have been successfully used in multiple studies over the years. These studies have also introduced incremental improvements to the methodologies when applying these methods to the research of online communication. However, one weakness when researching collaboration online is the relative lack of automation. Observing and establishing the communication context still requires manpower, though research in automated analysis is ongoing, especially in the field of social network analysis.*

## 4. DISCUSSION

The state of research progressed during the period investigated in this systematic mapping study. Fundamental tools for researching collaboration in CSCL were introduced and evolved over a series of papers. Likewise, empirical studies established the basic effectiveness of CSCL in the field of software engineering education, especially in student motivation, productivity and improved critical thinking. More recent publications have also considered online collaboration from the community point of view and presented case studies about successful global collaboration networks.

The results of a positive impact of CSCL on software engineering education match the more general literature reviews about CSCL in education. For example, Resta and Lafarrière [82] also conclude in their literature review that CSCL has been generally accepted to be beneficial to students in higher order thinking skills, student satisfaction and improved productivity.

Resta and Lafarrière [82] pointed out, however, that there were research gaps in the product variables of CSCL. In essence, they stated that while a lot research has been done about the effectiveness of CSCL, it still cannot be said which factors make CSCL the most efficient approach. While this systematic mapping study was not such an in-depth study as a literature review, it can be agreed that this research gap also existed in the field of CSCL in software engineering education at the time of publication. Also, the studies about wide online communities proved that they work in regard to learning efficiency, but more studies are still required to find out what exactly makes them successful.

### 4.1. Recommendations for Using Computer-Supported Collaborative Learning in Software Engineering Education

With a multitude of studies discussing the benefits of CSCL in software engineering education, adopting one of the approaches can be safely recommended for teachers to try. Because the great amount of different approaches, the main challenge is choosing a right approach for each course. For project and capstone – style courses industry standard collaboration tools like Github or Redmine can be recommended. These can be augmented by online tools that add mutual support methods for novice programmers, like an avenue for programming questions [22].

For more lecture and exercise class –type courses different approaches can work better. Methods that add positive interdependence, common tasks, collaborative behavior and student accountability work best [60]. Furthermore, adding interactive materials to a course using CSCL methods can increase student engagement and motivation [88]. According to literature adding group tasks to the course is not enough, but the assignment and the online environment have to be carefully designed to add positive interdependence for learning and individual responsibility.

### 4.2. Limitations

The main limitation is that a systematic mapping study is broader, but shallower than a full literature review. While this SMS can give outlines and directions for research in the chosen field of study, it cannot compare the research outcomes and details of methodology. Another limitation is the scope of the mapping study. The research articles were only considered when they handled the issue from the point of view of information technology and software engineering. Seven scientific databases were used for the article searches, and they should cover a great majority of the papers from the selected time period. One database, Google Scholar, had to be excluded because it did not provide all necessary search parameter options and filtering methods.

### 4.3. Future research

The field of computer-supported collaborative learning in software engineering and information technology has some areas where there are still research gaps. The basics of CSCL as a method and its benefits in a variety of settings have been established. However, CSCL is a broad topic and there are several different ranges and methods of implementing it in communities and classrooms. As Resta and Lafarrière [82] proposed, less emphasis should be put on comparing CSCL to other collaborative teaching methods, and different CSCL approaches should be compared to each other instead. This would make it possible to find the individual success factors that affect CSCL processes and environments.

The proposed research approach of more in-depth studies is already addressed, especially in more pedagogical publications, which were outside the scope of this mapping study [21, 33, 46, 49, 62]. With regard to the wider collaboration networks, which were one specific area of investigation in this mapping study, the process of researching specifics has not yet started, and papers are still establishing the basic beneficiality of globally connected CSCL environments. However, it should be noted that the articles about global CSCL are successfully establishing the fact and are providing a solid basis for future in-depth studies about narrower aspects of global CSCL.

## 5. CONCLUSION

Collaboration is increasingly in use in higher education and in software engineering education, with several new and novel applications coming up each year. Different avenues of collaboration are still being explored, but the importance of a pedagogical approach and organizational support, as well as the need of good, supporting software tools are already clear.

This article mapped the existing literature on computer-supported collaborative learning in software engineering education by searching articles in scientific publication databases. A total of 79 articles published between 2003 and 2013 were chosen on the basis of inclusion and exclusion criteria. The number of publications per year was between four and eight, with some outliers. The articles were arranged into categories based on the scope of collaboration and the main issue researched. The most numerous categories of articles were the ones introducing new CSCL tools (20 articles), and the most commonly used research approach was constructive research. A single class was the most common research setup size (20 articles). Several articles inspected a single community (10 articles) or multiple communities (11 articles).

The articles published in the field of science showed that CSCL is beneficial to learning in the fields of information technology and software engineering education, especially for student motivation, productivity and improved critical thinking. It was shown to work both in local and globally distributed communities. These cases and designs provided a large body of knowledge for implementing CSCL environments or scenarios, researching the occurrences of collaboration, and baselines for the design of new collaborative tools.

To sum up, the benefits of CSCL have been established. However, many studies up to this date still inspect single tools or general use cases. It is not yet clear which elements are the most essential ones for successful CSCL environments, and how global CSCL works compared to local environments. Future research would be best served by two separate approaches: studying individual CSCL elements closely, and developing and comparing large collaborative communities at the same time.

### ACKNOWLEDGEMENTS

We thank our colleagues Tommi Kähkönen and Andrey Maglyas for sharing their expertise on writing literature reviews in the field of software engineering.

*Information about the authors:*

**Antti Knutas** – MSc., Junior researcher at Lappeenranta University of Technology, department of Innovation and Software. His current research topics are computer supported collaborative learning, technology enhancing learning tools and learning analytics.

**Jouni Ikonen** – Associate Professor at Lappeenranta University of Technology, department of Innovation and Software. His current research interests include open access networks, data brokering between applications, technology enhancing learning tools and industry skills.

**Jari Porras** – Professor at Lappeenranta University of Technology, department of Innovation and Software. His research interests include technology enhanced learning, in particular group work and skill development.